\documentclass[fleqn,12pt]{article}
\usepackage{gc,amssymb}
\usepackage{graphics}
\usepackage[dvips]{graphicx}

\def\kappa{\varkappa}
\def\phii {\phi_{\infty}}
\def\phicr {\phi_{\rm cr}}

\def\mn{_{\mu\nu}}
\def\MN{^{\mu\nu}}
\def\mN{_\mu^\nu}

\def\M{{\mathbb M}}
\def\R{{\mathbb R}}
\def\S{{\mathbb S}}
\def\Z{{\mathbb Z}}

\def\cK{{\cal K}}
\def\tT{{\widetilde T}}

\def\oV{\overline{V}}
\def\oVinf{\overline{V}_{\infty}}
\def\vp{{\vec p}\,{}}

\def\Sch{Schwarzschild}
\def\dS{de Sitter}
\def\AdS{anti-de Sitter}
\def\RN{Reissner-Nordstr\"om}

\def\GR{general relativity}
\def\sph{spherically symmetric}
\def\ssph{static, spherically symmetric}

\def\bh{black hole}

\def\bw{brane world}

\heads{K.A. Bronnikov and B.E. Meierovich}
      {Gravitating global monopoles in extra dimensions and the brane
       world concept}

\begin{document}
\onecol
\thispagestyle{empty}  \rightline{\bf gr-qc/0507032}
\bigskip

\Title {Gravitating global monopoles in extra dimensions \yy
        and the brane world concept}

\Author{Kirill A. Bronnikov} {Center for Gravitation and Fundam.
        Metrology, VNIIMS, 3-1 M. Ulyanovoy St., Moscow 117313, Russia;\\
        Institute of Gravitation and Cosmology, PFUR,
        6 Miklukho-Maklaya St., Moscow 117198, Russia\\}

\Author{Boris E. Meierovich} {Kapitza Institute for Physical
        Problems, 2 Kosygina St., Moscow 117334, Russia\\
        Email: meierovich@yahoo.com; http://geocities.com/meierovich/}

\Abstract
  {Multidimensional configurations with Minkowski external space-time and a
   \sph\ global monopole in extra dimensions are discussed in the context
   of the brane world concept. The monopole is formed with a hedgehog-like
   set of scalar fields $\phi^i$ with a symmetry-breaking potential $V$
   depending on the magnitude $\phi^2 = \phi^i \phi^i$. All possible kinds
   of globally regular configurations are singled out without specifying the
   shape of $V(\phi)$. These variants are governed by the maximum value
   $\phi_m$ of the scalar field, characterizing the energy scale of symmetry
   breaking. If $\phi_m < \phicr$ (where $\phicr$ is a critical value of
   $\phi$ related to the multidimensional Planck scale), the monopole
   reaches infinite radii while in the ``strong field regime'', when
   $\phi_m\geq \phicr$, the monopole may end with a cylinder of finite
   radius or possess two regular centers. The warp factors of monopoles with
   both infinite and finite radii may either exponentially grow or tend to
   finite constant values far from the center. All such configurations are
   shown to be able to trap test scalar matter, in striking contrast to
   RS2 type five-dimensional models. The monopole structures obtained
   analytically are also found numerically for the Mexican hat potential
   with an additional parameter acting as a cosmological constant.  }

\section{Introduction}

   According to a presently popular idea, our observable Universe can be
   located on a four-dimensional surface, called the brane, embedded in a
   higher-dimensional manifold, called the bulk. This ``brane world''
   concept, suggested in the 80s \cite{Brane}, is broadly discussed
   nowadays, mainly in connection with the recent developments in
   supersymmetric string/M-theories \cite{M-theory}. A reason why we do not
   see any extra dimensions is that the observed matter is confined to the
   brane, and only gravity propagates in the whole bulk. There are numerous
   applications of the \bw\ concept to particle physics, astrophysics and
   cosmology, such as the hierarchy problem and the description of dark
   matter and dark energy \cite{bra-rev}.

   Most of the studies are restricted to infinitely thin branes with
   delta-like localization of matter. A well-known example is Randall and
   Sundrum's second model (RS2) \cite{RS2} in which a single Minkowski brane
   is embedded in a 5D anti-de Sitter (AdS) bulk.

   Thin branes can, however, be only treated as a rough approximation since
   any fundamental underlying theory, be it quantum gravity, string or
   M-theory, must contain a fundamental length beyond which a classical
   space-time description is impossible. It is therefore necessary to
   justify the infinitely thin brane approximation as a well-defined limit
   of a smooth structure, a thick brane, obtainable as a solution to coupled
   gravitational and matter field equations. Such a configuration is then
   required to be globally regular, stable and properly concentrated around
   a 3D surface which is meant to describe the observed spatial dimensions.
   Topological defects, emerging in phase transitions with spontaneous
   symmetry breaking (SSB), are probably the best candidates for this role.

   It should be mentioned that the evolution of the Universe, according to
   modern views, contained a sequence of phase transitions with SSB. A
   decisive step toward cosmological applications of the SSB concept was
   made in 1972 by Kirzhnits \cite{Kirzhnits}. He assumed that, as in the
   case of solid substances, a symmetry of a field system, existing at
   sufficiently high temperatures, could be spontaneously broken as the
   temperature falls down. A necessary consequence of such phase transitions
   is the appearance of topological defects. The first quantitative analysis
   of the cosmological consequences of SSB was given by Zel'dovich, Kobzarev
   and Okun' \cite{Zeldovich et al}. Later on, the SSB phenomenon and
   various topological defects were widely used in inflationary Universe
   models and in attempts to explain the origin of the large-scale structure
   of the Universe, see, e.g., \cite{inflation, Vilenkin Shellard}.

   The properties of global topological defects are generally described with
   the aid of a multiplet of scalar fields playing the role of an order
   parameter. If a defect is to be interpreted as a brane world, its
   structure is determined by the self-gravity of the scalar field system
   and may be described by a set of Einstein and scalar equations.

   In this paper we analyze the gravitational properties of candidate (thick)
   brane worlds with 4D Minkowski metric as global topological defects in
   extra dimensions. Our general formulation covers such particular cases as
   a brane (domain wall) in 5D space-time (one extra dimension), a global
   cosmic string with winding number $n=1$ (two extra dimensions), and
   global monopoles (three or more extra dimensions). We restrict ourselves
   to Minkowski branes since most of the existing problems are clearly seen
   even in these comparatively simple systems; on the other hand, in the
   majority of physical situations, the inner curvature of the brane itself
   is much smaller than the curvature related to brane formation, therefore
   the main qualitative features of Minkowski branes should survive in
   curved branes.

   Brane worlds as thick domain walls in a 5D bulk have been discussed in
   many papers (see, e.g., \cite{5D} and references therein). Such systems
   were analyzed in a general form in Refs.\,\cite{thick2, thick3}, without
   specifying the symmetry-breaking potential; it was shown, in particular,
   that all regular configurations should have an AdS asymptotic. So all
   possible thick branes are merely regularized versions of the RS2 model,
   with all concomitant difficulties in material field confinement.
   Thus, it has been demonstrated \cite{thick3} that a test scalar field has
   a divergent stress-energy tensor (SET) infinitely far from the brane, at
   the AdS horizon. A reason for that is the repulsive gravity of RS2 and
   similar models: gravity repels matter from the brane and pushes it
   towards the AdS horizon. To overcome this difficuty, it is natural to try
   a greater number of extra dimensions. This was one of the reasons for us
   to consider higher-dimensional bulks.

   We study the simplest possible realization of this idea, assuming a
   \ssph\ configuration of the extra dimensions and a thick Minkowski brane
   as a concentration of the scalar field SET near the center. The possible
   trapping properties of gravity for test matter are then determined by the
   behavior of the so-called warp factor (the metric coefficient acting as a
   gravitational potential) far from the center, and we indeed find classes
   of regular solutions where gravity is attracting.

   Some of our results repeat those obtained in Refs.\,\cite{vil, shap}
   which have discussed global and gauge (`t Hooft-Polyakov type) monopoles
   in extra dimensions; see a more detailed comparison in \sect 7.

   The paper is organized as follows. In \sect 2 we formulate the problem,
   introduce space-times with global topological defects in the extra
   dimensions, write down the equations and boundary conditions and
   demonstrate a connection between the possibility of SSB and the
   properties of the potential at a regular center.

   In \sect 3 we briefly discuss the trapping problem for RS2 type
   domain-wall models and show that they always possess repulsive gravity
   and are unable to trap matter in the form of a test scalar field.

   \sect 4 is devoted to a search for regular global monopole solutions in
   higher dimensions by analyzing their asymptotic properties far from the
   center. All regular configurations are classified by the behavior of the
   spherical radius $r$ and by the properties of the potential. This leads
   to separation of ``weak gravity'' and ``strong gravity'' regimes, related
   to maximum values of the scalar field magnitude.

   In the weak gravity regime, the spherical radius $r$ tends to infinity
   along with the distance from the center. Such moderately curved
   configurations exist without any restrictions of fine-tuning type.

   If the scalar field magnitude exceeds some critical value, the radius
   $r$ either tends to a finite value far from the center or returns to zero
   at a finite distance from the center, thus forming one more center which
   should also be regular. Some cases require fine tuning of the parameters
   of the potential, so one may believe that static configurations can only
   exist if the scalar and gravitational forces are somewhat mutually
   balanced.

   In \sect 5 we show that, in contrast to domain walls, global monopoles in
   different regimes do provide scalar field trapping on the brane.

   \sect 6 is a brief description of numerical experiments with the Mexican
   hat potential admitting shifts up and down, equivalent to introducing a
   bulk cosmological constant. Their results confirm and illustrate the
   conclusions of \sect 4.

   \sect 7 summarizes the results.

\section{Problem setting}

\subsection{Geometry}

   We consider a $(D = d_0+d_1+1)$-dimensional space-time with the structure
   $\M^{d_0} \times \R_u \times \S^{d_1}$ and the metric
\beq                                                         \label{ds}
     ds^{2} = \e^{2\gamma (u)}\eta\mn  dx^{\mu} dx^{\nu}
            -\left( \e^{2\alpha
                (u)}du^{2} + \e^{2\beta (u)}d\Omega^{2} \right) .
\eeq
   Here $\eta\mn dx^\mu dx^\nu = dt^2 - (d \vec x)^2$ is the Minkowski metric
   in the subspace $\M^{d_0}$, $\eta\mn = \diag ( 1,\ -1,...,\ -1)$;
   $d\Omega $ is a linear element on a $d_{1}$-dimensional unit sphere
   $\S^{d_1}$; $\alpha$, $\beta$ and $\gamma$ are functions of the radial
   coordinate $u$ with the definition domain $\R_u \subseteq \R$, to be
   specified later. The Riemann tensor has a diagonal form, and its nonzero
   components are
\bear                                                          \label{Riem}
    R_{\ \ \rho \sigma }^{\mu \nu } \eql -\e^{-2\alpha }\gamma'^2
            \delta _{\ \ \rho \sigma }^{\mu \nu },
\nn
    R_{\ \ cd}^{ab} \eql \left( \e^{-2\beta }
            -\e^{-2\alpha }\beta'^2 \right)\delta _{\ \ cd}^{ab},
\nn
    R_{\ \ u\nu }^{u\mu } \eql -\delta _{\nu }^{\mu }\e^{-\gamma -\alpha}
            \left( \e^{\gamma -\alpha }\gamma'\right)',
\nn
    R_{\ \ ub}^{ua} \eql  -\delta _{b}^{a}\e^{-\beta -\alpha }
            \left( \e^{\beta -\alpha }\beta'\right)',
\nn
    R_{\ \ b\nu }^{a\mu } \eql -\delta_{\nu }^{\mu }
            \delta _{b}^{a}\e^{-2\alpha }\gamma'\beta'.
\ear
Here
\beq
     \delta\MN{}_{\rho \sigma}
    =\delta _{\rho }^{\mu }\delta _{\sigma }^{\nu }
            -\delta_{\sigma}^{\mu }\delta_{\rho }^{\nu },
                                            \label{Delta*munu_rosigma}
\eeq
    and similarly for $\delta_{\ \ cd}^{ab}$. Greek indices $\mu,\nu,...$
    correspond to $d_{0}$-dimensional space-time, and Latin indices
    $a,b,...$ to $d_{1}$ angular coordinates on S$^{d_{1}}$. We mostly bear
    in mind the usual dimension $d_0 = 4$ but keep $d_0$ arbitrary for
    generality.

    A necessary condition of regularity is finiteness of the Kretschmann
    scalar $\cK = R_{\ \ CD}^{AB}R_{\ \ AB}^{CD}$. (Capital indices $A,B,...$
    correspond to all $D$ coordinates.) In our case, $\cK$ is a sum of
    squares of all nonzero $R_{\ \ CD}^{AB}$. Hence, in regular
    configurations all components of the Riemann tensor (\ref{Riem}) are
    finite.

    For the Ricci tensor we have
\bear                                                      \label{Ricci}
    R_{\mu }^{\nu } \eql
    -\delta\mN  \e^{-2\alpha }\left[ \gamma''
    + \gamma'(-\alpha'+ d_0 \gamma' + d_{1}\beta )\right],
\nn
    R_{u}^{u} \eql
    -\e^{-2\alpha }\left[ d_0 (\gamma''+\gamma'{}^2 - \alpha'\gamma')
    + d_{1}(\beta''+ \beta'{}^2 - \alpha'\beta')\right]
\nn
    R_{m}^{n} \eql \e^{-2\beta }(d_1 - 1)\delta _{m}^{n}
        -\delta_{m}^{n}\e^{-2\alpha }
    \left[ \beta''+\beta'(-\alpha'+d_{0}\gamma'+ d_{1}\beta')\right].
\ear

\subsection{Topological defects}

   A global defect with a nonzero topological charge can be constructed as a
   multiplet of $d_1+1$ real scalar fields $\phi^k$, in the same way as,
   e.g., in $\cite {Bronnikov Multi-D}$. It comprises a hedgehog
   configuration in $\R_u \times \S^{d_1}$:
\[
    \phi^k = \phi (u) n^k (x^a),
\]
   $n^k$ is a unit vector in the $d_1+1$-dimensional Euclidean target space
   of the scalar fields: $n^k n^k =1$.

   The total Lagrangian of the system is taken in the form
\beq                                                         \label{Lagr}
    L = \frac{R}{2\kappa^2}
            + \Half g^{AB}\d_A\phi^k \d_B \phi^k - V (\phi),
\eeq
   where $R$ is the $D$-dimensional scalar curvature, $\kappa^2 $ is the
   $D$-dimensional gravitational constant, and $V$ is a symmetry-breaking
   potential depending on $\phi^2 (u) = \phi^{a} \phi^{a}$.

   In case $d_1 = 0$ there is only one extra dimension. The topological
   defect is a flat domain wall. Combined with $d_0=4$, it is widely
   considered with reference to our Universe. Regular thick Minkowski branes
   supported by scalar fields with arbitrary potentials were analyzed in
   \cite{thick2, thick3}, see also \sect 3 of the present paper.

   The case $d_1=1$ is a global cosmic string with the winding number $n=1$.
   If $d_0=2$, it is a cosmic string in four dimensions, whose gravitational
   properties are reviewed in \cite{Meierovich Uspehi}.
   The case $d_0 = 4$ corresponds to a string in extra dimensions.

   $d_1=2$, $d_0=1$ is the case of a global monopole in our 4-dimensional
   space-time. We have analyzed it in detail in \cite{Bronnikov Global mon}.
   The case $d_1 > 2$, $d_0 = 1$ is its multidimensional generalization to
   \ssph\ space-times with $d_1$-dimensional rather than 2-dimensional
   coordinate spheres \cite{Bronnikov Multi-D}. It was shown that such a
   heavy multidimensional global monopole leads to a multidimensional
   cosmology where the symmetry-breaking potential at late times can mimic
   both dark matter and dark energy.

   In case $d_0 = 4$, $d_1 > 2$ we have a multidimensional global monopole
   entirely in the extra space-like dimensions. Different models of this
   kind were studied in Refs.\,\cite{vil, shap, Benson Cho, Cho and
   Vilenkin}. In particular, such a monopole in extra dimensions was used in
   an attempt to explain the origin of inflation \cite{Cho and Vilenkin}.

\subsection{Field equations}

    We are using the Einstein equations in the form
\[
     R_{A}^{B} = -\kappa^2 \tT_{A}^{B},\cm
        \tT_{A}^{B} = T_{A}^{B}-\frac{\delta_{A}^{B}}{D-2} T^C_C,
\]
    where $T_A^B$ is the stress-energy tensor of the scalar field
    multiplet. For our hedgehog configuration,
\bear
    \tT\mN  \eql - 2V \delta\mN /(D-2),
\nn
    \tT_u^u \eql - 2V/(D-2) - \e^{-2\alpha }\phi'^2,
\nn
    \tT_a^b \eql - 2V \delta_a^b /(D-2) - \e^{-2\beta}\delta_a^b \phi^2.
\earn

    So far we did not specify the radial coordinate $u$. For our purposes,
    the most helpful is the {\bf Gaussian} gauge such that the real distance
    $l$ along the radial direction is taken as a coordinate:
\beq
    u \equiv l, \cm \alpha \equiv 0,                          \label{Gauss}
\eeq
    and the metric is
\beq                                                          \label{ds-g}
     ds^{2} = \e^{2\gamma (l)} \eta\mn  dx^{\mu }dx^{\nu }
                    -\left(dl^{2} + \e^{2\beta (l)}d\Omega^{2}\right).
\eeq
    Then two independent components of the Einstein equations take
    the following form (the prime now denotes $d/dl$):
\bearr
    \gamma'' + d_0 \gamma'^2 + d_1 \beta'\gamma'
            = -\frac{2\kappa^2}{D-2}  V,                      \label{00-g}
\\[2pt] \lal
      \beta'' + d_0 \beta'\gamma' + d_1 \beta'^2
        = (d_1 - 1 - \kappa^2 \phi^2)\e^{-2\beta}
            -\frac{2\kappa^2 }{D-2} V.                        \label{22-g}
\ear
    The Einstein equation $G_l^l = -\kappa^2 T^l_l$ (where $G_A^B$ is the
    Einstein tensor) is free of second-order derivatives:
\bearr
    (d_0 \gamma' + d_1 \beta')^2                             \label{int-g}
                    -d_0 \gamma'^2 - d_1 \beta'^2
    = \kappa^2 (\phi'^2 - 2V)
         + d_1 \e^{-2\beta}  (d_1 - 1 - \kappa^2 \phi^2).
\ear
    The scalar field equations
\[
      \nabla^A \nabla_A \phi^k + \d V/\d\phi^k =0
\]
    combine to yield an equation for $\phi(l)$:
\beq
    \phi'' + (d_0\gamma' + d_1\beta') \phi' - d_1 \e^{-2\beta}\phi
        = \frac{d V}{d \phi}.                                \label{phi-g}
\eeq
    Due to the Bianchi identities, it is a consequence of the Einstein
    equations (\ref{00-g})--(\ref{int-g}). On the other hand, (\ref{int-g})
    is a first integral of \eqs (\ref{00-g}), (\ref{22-g}) and (\ref{phi-g}).

    In our analytical study, we do not specify any particular form of
    $V(\phi)$. We, however, suppose that $V$ has a maximum at $\phi = 0$ and
    a minimum at some $\phi =\eta > 0$, so that $V'(0) =V'(\eta)=0$.
    For convenience, we do not single out a cosmological constant which may
    be identified with a constant component of the potential $V$ or, in
    many cases, with its minimum value.

    The parameter $\eta$ (as the scalar field itself) has the dimension
    $[l^{-(D-2)/2}]$ and thus specifies a certain length scale
    $\eta^{-2/(D-2)}$ and energy scale $\eta^{2/(D-2)}$ (we are using the
    natural units, such that $c=\hbar =1$.) In the conventional case $D=4$,
    $\eta$ has the dimension of energy and characterizes the SSB energy
    scale.

\subsection {Regularity conditions. A regular center}

    For the geometry to be regular, we should require finite values of all
    Riemann tensor components (\ref{Riem}). In the Gaussian gauge
    (\ref{Gauss}) the regularity conditions are as simple as that
\beq                              \label{reg-g}
       \beta',\ \ \ \beta'', \ \ \ \gamma', \ \ \ \gamma''\ \ \
                \mbox{should be finite}.
\eeq

    For $d_1 > 0$, in addition to (\ref{reg-g}), a special regularity
    condition is needed at a center, which is a singular point of the
    spherical coordinates in $\R_u \times S^{d_1}$. A center is a point
    where the radius $r \equiv \e^{\beta}$ turns to zero, The regularity
    conditions there, also following from finiteness of the Riemann tensor
    components (\ref{Riem}), are the same as in a usual \ssph\ space-time:
    in terms of an arbitrary $u$ coordinate, they read
\bear                                                     \label{reg_c}
      \gamma = \gamma_c + O(r^2), \cm
      \e^{\beta-\alpha} |\beta'| = 1 + O(r^2) \cm {\rm as}\quad r\to 0,
\ear
    where $\gamma_c$ is a constant which can be set to zero by a proper
    choice of scales of the coordinates $x^\mu$. The second condition in
    (\ref{reg_c}) follows, for $d_1 > 1$, from finiteness of the Riemann
    tensor components $R^{ab}{}_{cd}$, see (\ref{Riem}). Its geometric
    meaning is local Euclidity at $r=0$ that implies $dr^2 = dl^2$, i.e., a
    correct circumference to radius ratio for small circles. In the special
    case $d_1=1$, so that the factor space $\R_u \times S^{d_1}$ is
    two-dimensional, we evidently have $R^{ab}{}_{cd} \equiv 0$, but the
    second condition in (\ref{reg_c}) should still be imposed to avoid a
    conical singularity.

    It is natural to put $l=0$ at a regular center, so that $l$ is the
    distance from the center.

    Regularity of the Ricci tensor components $R^A_B = R^{AC}{}_{BC}$
    implies regularity of the stress-energy tensor $T_A^B$, whence it
    follows that, at any regular point and with any radial coordinate,
\beq
      |V| < \infty,   \cm                                   \label{reg_Tmn}
      \e^{-\beta}|\phi| < \infty,   \cm
      \e^{-\alpha}|\phi'| < \infty.
\eeq

\subsection{Boundary conditions}

    {\bf Domain walls.}
    In case $d_1 = 0$, the metric (\ref{ds}) or (\ref{ds-g}) describes a
    plane-symmetric 5D space-time, the coordinate $l$ ranges over the
    whole real axis, and the broken symmetry is $\Z_2$, mirror symmetry with
    respect to the plane $l=0$. The topological defect is a domain wall
    separating two vacua, corresponding to two values of a single real
    scalar field $\phi$, say, $\phi = \pm \eta$. Accordingly, we assume that
    $\phi (l)$ is an odd function whereas $\gamma(l)$ and $V(\phi)$ are even
    functions, and the conditions at $l=0$ are
\beq
     \gamma(0) = \gamma'(0) = \phi (0) =0.                 \label{ini-5}
\eeq
    We thus have three initial conditions for the third-order set of
    equations (\ref{00-g}), (\ref{int-g}) [\eq (\ref{phi-g}) is their
    consequence] since in this case the unknown function $\beta$ is absent.

\medskip\noi
    {\bf Global strings and monopoles.} For $d_1 >0$,
    the regular center requirement leads to the following boundary
    conditions for \eqs (\ref{00-g})--(\ref{int-g}) at $l=0$:
\beq                                                        \label{cond_0}
     \phi (0) = \gamma (0) = \gamma '(0) = r(0) = 0,\cm   r'(0) =1.
\eeq

    We have five initial conditions for a fifth-order set of equations.
    However, $l = 0$, being a singular point of the spherical coordinate
    system (not to be confused with a space-time curvature singularity), is
    also a singular point of our set of equations. As a result, the
    requirements of the theorem on the solution existence and uniqueness
    for our set of ordinary differential equations are violated. It turns
    out that the derivative $\phi'(0)$ remains undetermined by
    (\ref{cond_0}). If we set $\phi'(0) =0$, we obtain a trivial (symmetric)
    solution with $\phi \equiv 0$ and a configuration without a topological
    defect. In case $V(0) = 0$ we arrive at the flat $D$-dimensional metric:
    we have then $\gamma\equiv 0$ and $r\equiv l$ in (\ref{ds-g}). If,
    however, $V(0) \ne 0$, the corresponding exact solutions to the Einstein
    equations for $d_0 >1$, $d_1 > 1$ are yet to be found. A direct
    inspection shows that it cannot be de Sitter or AdS space: the constant
    curvature metrics are {\it not\/} solutions to the vacuum Einstein
    equations with a cosmological constant.

    Nontrivial solutions take place if $\phi'(0) \neq 0$ and can correspond
    to SSB. One can notice that the very possibility of SSB appears as a
    result of violation of the solution uniqueness at $r=0$ provided that a
    maximum of the potential $V(\phi)$ at $\phi =0$ corresponds to the
    center. The lacking boundary condition which may lead to a unique
    solution can now follow from the requirement of regularity at the other
    extreme of the range of $l$ whose nature is in turn determined by the
    shape of the potential.

    In what follows, assuming a regular center, we will try to
    find all possible conditions at the other extreme of the range $\R_l$
    of the Gaussian radial coordinate, providing the existence of globally
    regular models with the metric (\ref{ds-g}). In other words, we seek
    solutions with such asymptotics that the quantities (\ref{Riem}) are
    finite. All other regularity conditions, such as (\ref{reg_Tmn}), then
    follow.

    In doing so, we will not restrict in advance the possible shape of the
    potential $V(\phi)$. The cases under consideration will be classified by
    the final values of $r=\e^\beta$ (infinite, finite or zero) and $V$
    (positive, negative or zero). The scalar field $\phi$ is assumed to be
    finite everywhere.

    Without loss of generality, we assume $\phi'(0) > 0$ near $l=0$, i.e.,
    that $\phi$ increases, at least initially, as we recede from the center.

\section {Domain walls and the problem of matter confinement}

    Below we will mostly consider configurations with $d_1\geq 2$
    which correspond to a global monopole in the \sph\ space $\R_u\times
    \S^{d_1}$. Before that, let us briefly discuss the problem of matter
    confinement on the brane and its difficulty in the 5-dimensional case.

    The metric coefficient $\e^{2\gamma}$ in (\ref{ds}), sometimes called
    the warp factor, actually plays the role of a gravitational potential
    that determines an attractive or repulsive nature of gravity with
    respect to the brane. If it forms a potential well with a bottom on (or
    very near) the brane, there is a hope that matter, at least its
    low-energy modes, will be trapped.

    It has been shown, in particular, that spin-1/2 fields are localized due
    to an increasing warp factor in (1+4)- and (1+5)-dimensional models
    \cite{bajk, oda}. It was also repeatedly claimed that, in (1+4)
    dimensions, a brane with an exponentially descreasing warp factor (as,
    e.g., in the RS2 model) can trap spin 0 and 2 fields. Our calculation
    for a scalar field shows that it is not the case.

    A gravitational trapping mechanism suggested in Refs.\,\cite{gog} was
    characterized there as a universal one, suitable for all fields. It is
    based on non-exponential warp factors, which increase from the brane
    and approach finite values at infinity. This mechanism was exemplified
    in \cite{gog-sing} with a special choice of two so-called ``smooth
    source functions'' in the SET, describing a continuous distribution of
    some phenomenological matter and vanishing outside the brane.

    Our analysis uses more natural assumptions: a scalar field system
    admitting SSB, without any special choice of the symmetry breaking
    potential, under the requirement of global regularity.

    Let us briefly show, following Refs.\,\cite{thick2, thick3} (but in
    other coordinates) that this approach in (4+1) dimensions always leads
    to a decaying warp factor whatever be the choice of $V(\phi)$ and that
    such a system cannot trap a test scalar field. So consider
    a domain wall in 5 dimensions, so that $l \in \R$, in our equations we
    put $d_1 = 0$, the unknown $\beta (l)$ is absent, while \eqs
    (\ref{00-g}) and (\ref{phi-g}) for $\gamma$ and the single scalar field
    $\phi$ read
\bearr                                                 \label{00-5}
    \gamma'' + d_{0}\gamma'^2 = -\frac{2\kappa^{2}}{d_{0}-1} V,
\\ \lal
    \phi'' + d_{0} \gamma'\phi' - \frac{d V}{d \phi } =0.      \label{phi-5}
\ear
     Their first integral (\ref{int-g}) reduces to
\beq
    \gamma'{}^2 = -\frac{\kappa^{2}}{d_0(d_0-1)} (2V - \phi'{}^2) ,
                                                 \label{int-5}
\eeq
    The initial conditions at $l=0$ corresponding to $\Z_2$ symmetry
    (broken for the scalar field but preserbed for the geometry) have
    the form (\ref{ini-5}).

    Excluding $V$ from (\ref{00-5}) and (\ref{int-5}) and integrating
    subject to (\ref{ini-5}), we obtain
\beq                                                       \label{gam'-5}
     (d_0 - 1)\gamma'(l) = - \kappa^2 \int_{0}^{l}\phi'{}^2 dl,
\eeq
    and we conclude that $\gamma'(l)$ is negative at all $l > 0$ and
    describes gravitational repulsion from the brane; moreover,
    $\e^{-\gamma}$ monotonically grows with growing $l$. The only possible
    regular solution corresponds to $|\gamma'(\infty)| < \infty$. Since in
    this case $\gamma''(\infty) = 0$, it follows from \eq (\ref{00-5}) that
    $V(\infty) < 0$, corresponding to a negative cosmological constant
    $\Lambda = \kappa^2 V(\infty)$. So the only possible regular asymptotic
    is AdS, with
\beq
      \e^\gamma \approx a\e^{-hl}, \cm a,\,h= \const,
            \cm  h = \sqrt{-\Lambda/6}.
\eeq
    The constant $a$ depends on the particular shape of $V(\phi)$. At $l =
    \infty$, there is an AdS horizon ($\e^{\gamma}=0$), which, like a black
    hole horizon, attracts matter and prevents its trapping by the brane.

    Let us show this for $d_0 = 4$ and a test scalar field $\chi$
    with the Lagrangian
\beq                                                           \label{L_hi}
     L_{\chi} = \half \d_A \chi^* \d^A \chi - \half m_0^2\chi^* \chi
                - \half \lambda \phi^2 \chi^*\chi,
\eeq
    where $\chi^*$ is the complex congugate field, and the last term
    describes a possible interaction between $\chi$ and the wall scalar
    field $\phi$; $\lambda$ is a coupling constant. The field $\chi (x^{A})$
    satisfies the linear homogeneous (modified Fock-Klein-Gordon) equation
\beq                                                        \label{KG}
    \frac{1}{\sqrt{g}}\d_{A}\left( \sqrt{g}g^{AB}
                    \d_{B}\chi \right) + (\lambda \phi^2 + m_0^2) \chi =0.
\eeq
    Its coefficients depend on $l$ only, and
    $\chi (x^{A})$ may be sought for in the form
\beq                                                            \label{sepa}
    \chi (x^{A}) = X(l) \exp (-ip_{\mu} x^{\mu}),
                            \qquad \mu =0,1,2,3.
\eeq
    where $p_\mu = (E,\ \vec p)$ is a constant 4-momentum.
    The function $X(l)$ determines the $\chi$ field distribution across
    the brane and satisfies the equation
\beq                                                        \label{X''}
    X'' + 4\gamma' X' + \left[ \e^{-2\gamma} (E^2-\vp^2)
                      - \lambda \phi^2 -m_0^2)\right] X = 0.
\eeq

    The $\chi$ field is able to describe particles localized on the brane
    only if its SET, $T\mN [\chi]$, is finite in the whole 5-space and
    decays sufficiently rapidly at large $l$. As an evident necessary
    condition of localization, the $\chi$ field energy per unit 3-volume of
    the brane should be finite, i.e.,
\bearr                                                       \label{E_tot}
     E_{\rm tot}[\chi] = \int_{-\infty}^{\infty} T^t_t \sqrt{g} dl
     = \int_{0}^{\infty}
            \e^{4\gamma}\Bigl[ \e^{-2\gamma}(E^2 + \vp^2)X^2
                  + (m_0^2 +\lambda\phi^2)X^2 + X'^2 \Bigr] dl < \infty.
\ear
    The inequality (\ref{E_tot}) implies a finite norm of the
    $\chi$ field defined as
\beq                                 \label{hi-norm}
    \|\chi\|^2 = \int_{-\infty }^{\infty }\sqrt{g}\,\chi^*\chi\,dl
               = \int_{-\infty }^{\infty } \e^{4\gamma}\,X^2 \,dl.
\eeq

    At large $l$, since $\e^{-2\gamma}\to \infty$, the terms with $\lambda$
    and $m_0$ in \eq (\ref{X''}) may be neglected, and the equation
    determining the behavior of $\chi$ at large $l$ may be written as
\bearr                                                   \label{X''as}
     X'' -4h\,X' + P^2 \e^{2hl}\, X =0, \cm P^2 = \frac{E^2-\vp^2}{a^2 h^2}.
\ear
    It is solved in terms of Bessel functions, and the solution has the
    asymptotic form
\beq                                                        \label{hi-as}
      X = C\e^{3hl/2} \sin (P\e^{hl} + \varphi_0 ),  \cm
            l\to \infty,
\eeq
    where $C$ and $\varphi_0$ are integration constants. We see that the
    magnitude (\ref{hi-as}) not only does not vanish as $l\to \infty $
    but oscillates with an increasing amplitude. As a result, the SET
    components $T\mN [\chi]$ are infinite at $l = \infty$. Moreover, the
    integral (\ref{E_tot}) behaves as $\int\e^{hl} dl$ and diverges. Though,
    the normalization integral (\ref{hi-norm}) converges since the integrand
    behaves as $\e^{-hl}$. The latter result is sometimes treated as a
    sufficient condition for localization, but, in our view, it is not true
    since the very existence of the brane configuration is put to doubt if
    the test field SET is somewhere infinite.

    Thus a test scalar field with any mass tends to infinity as $l\to
    \infty$ and develops an infinite SET; even its interaction with the
    $\phi$ field that supports the brane does not improve the situation. We
    conclude that a single extra dimension is insufficient for providing
    gravitational attraction of matter to a regular isolated brane.

\section {A search for regular asymptotics}

    Let us now consider the field equations (\ref{00-g})--(\ref{phi-g}) for
    global monopoles, assuming $d_1\geq 2$. The string case $d_1=1$ is left
    aside since it has some peculiarities which need a special study.

\subsection* {A: Solutions with the asymptotic $r\to\infty$}

    Let us denote
\beq                                                      \label{def-oV}
     \oV = \frac{2\kappa^2 V}{D-2}, \cm
                                \oVinf = \oV\Big|_{r\to\infty}.
\eeq

    Evidently, $l\to\infty$ as $r\to\infty$ since otherwise we would have
    $\beta'\to\infty$, violating the regularity conditions. The derivatives
    $\beta'$ and $\gamma'$ should tend to certain constant values, to be
    denoted $\beta'_\infty$ and $\gamma'_\infty$, respectively. Both
    $\beta''$ and $\gamma''$ vanish as $l\to \infty$. Moreover, in \eq
    (\ref{22-g}), the second term of the right-hand side also vanishes.
    Therefore, in the leading order of magnitude, \eqs (\ref{00-g}) and
    (\ref{22-g}) take the form
\bear
    \gamma'_\infty (d_0 \gamma'_\infty + d_1 \beta'_\infty) \eql - \oVinf,
\nn
    \beta'_\infty (d_0 \gamma'_\infty + d_1 \beta'_\infty) \eql - \oVinf.                   \label{A}
\ear
    Consider separately the cases $\oVinf \ne 0$ and $\oVinf =0$.

\subsubsection*{A1: $\oVinf \ne 0$}

    \eqs (\ref{A}) immediately give
\beq                                                        \label{bc_A1}
    \beta'_\infty = \gamma'_\infty = \sqrt{-\oVinf/(D-1)}, \cm \oVinf < 0.
\eeq
    An evident necessary condition of the existence of regular configurations
    is $\oVinf \leq 0$. We thus obtain $\e^{\beta} \sim \e^{\gamma} \sim
    \e^{\beta'_\infty l}$, and the metric takes the asymptotic form
\beq                                                       \label{ds_A1}
      ds^2 \approx C_1 \e^{2 \beta'_\infty l} \eta\mn dx^{\mu} dx^{\nu}
            - dl^2
                - C_2 \e^{2 \beta'_\infty l} d\Omega^2,
\eeq
    with some positive constants $C_1$ and $C_2$.
    \eq (\ref{int-g}) holds automatically provided $\phi'(\infty) =0$, as
    should be the case if we assume a finite asymptotic value of $\phi$.
    Finally, in \eq (\ref{phi-g}) all terms but $dV/d\phi$ manifestly vanish
    as $l\to\infty$, hence $dV/d\phi$ vanishes as well, which should be the
    case if the field $\phi$ reaching an extremum of the potential $V$.

    The condition of finiteness of $\phi$ as $l \to \infty$ separates a
    family of regular solutions among the continuum of integral curves
    leaving the regular center with different slopes $\phi'(0)$. As is
    confirmed by numerical experiments, if the potential has only one
    extremum (minimum) $\oVinf < 0$, then there can be only one regular
    solution with $r \to \infty$, $l \to \infty$. However, there can be
    numerous regular solutions if the potential has several extremum points
    $\oVinf < 0$.

    In particular, if the initial maximum of the potential is located
    below the zero level, $V(0)\leq 0$, then there can be a continuum of
    regular integral curves starting from the regular center and returning
    to $\phi=0$ at $l \to \infty$. As can be verified numerically (see
    \sect 4), there is a bunch of such curves parameterized by $\phi'(0) \in
    (0,\ \phi'_s)$, where $\phi'(0)=\phi'_s$ corresponds to a limiting
    regular curve (separatrix), also starting at $\phi(0) = 0$ but ending
    at the minimum $V(\eta)$.

    The metric (\ref{ds_A1}) solves the Einstein equations with the
    stress-energy tensor $T_A^B = \delta_A^B V_{\infty}$, having the
    structure of a (negative) cosmological term. Moreover, according to
    (\ref{Riem}), the Riemann tensor has the structure of a constant
    curvature space at large $l$. In other words, such solutions have an
    anti-de Sitter (AdS$_D$) asymptotic far from the center. However, the
    metric (\ref{ds_A1}) is not a solution to our equations in the whole
    space even in case $\phi = \const$. As was already mentioned, for
    $d_0 > 1$ and $d_1 > 1$, the constant curvature metrics (dS$_D$ and
    AdS$_D$) are {\it not\/} solutions to the vacuum Einstein equations with
    a cosmological constant.

\subsubsection*{A2: $\oVinf = 0$}

    \eqs (\ref{A}) are solved either by $\beta'_\infty = \gamma'_\infty =0$
    or by $d_0 \gamma'_\infty + d_1 \beta'_\infty =0$.  When, however, we
    substitute the second condition to \eq (\ref{int-g}), taking into
    account that $\phi'\to 0$ at large $l$, we get $d_0 {\gamma'_\infty}^2 +
    d_1 {\beta'_\infty}^2 = 0$ and return to $\beta'_\infty = \gamma'_\infty
    = 0$. Thus both $\beta'$ and $\gamma'$ vanish at infinity, and we can
    try to seek them as expansions in inverse powers of $l$:
\beq                                                             \label{l-}
      \beta'  = \frac{\beta_1}{l} + \frac{\beta_2}{l^2} + \ldots,
  \cm
      \gamma' = \frac{\gamma_1}{l} + \frac{\gamma_2}{l^2} + \ldots,
\eeq
    Then $O(l^{-2})$ is the senior order in the Einstein equations, and, to
    avoid a contradiction, $r^{-2} = \e^{-2\beta}$ should be of the order
    $O(l^{-2})$ or smaller. Moreover, since we assume that $\phi$ tends to a
    finite value $\phii >0$, we have $\phi' = o(1/l)$, and the scalar field
    equation (\ref{phi-g}) shows that $dV/d\phi = O(l^{-2})$ or smaller,
    i.e., $\phii$ should be an extremum of $V(\phi)$. If $\phi(l)$ grows
    monotonically to $\phii >0$, then $\phii$ is a minimum of $V$ since
    according to (\ref{phi-g}) $dV/d\phi < 0 $ as $\phi \to \phii$.
    However, if $V(0)=0$, one cannot exclude that $\phi$ returns to zero as
    $l \to \infty$, see item (c) below.

    In case $\phi \to \phii >0$, since $V_{\infty} = dV/d\phi (\phii) =0$,
    $V(\phi)$ is decomposed as
\beq
     V(\phi) = \Half V_{\phi\phi} (\phii)                      \label{V-A2}
                        \cdot(\phi - \phii)^2 + \ldots,
\eeq
    (where $V_{\phi\phi} = d^2 V/d\phi^2$), so that $V = o(l^{-2})$.
    As a result, \eqs (\ref{00-g})--(\ref{int-g}) with (\ref{l-}) lead to
\bear                                                         \label{00-A2}
    \gamma_1 (-1 + d_0 \gamma_1 + d_1 \beta_1) \eql 0,
\\                                                            \label{22-A2}
     \beta_1 (-1 + d_0 \gamma_1 + d_1\beta_1) \eql
                \frac{l^2}{r^2}(d_1-1-\kappa^2\phii^2),
\\                                                            \label{int-A2}
    (d_0\gamma_1 + d_1\beta_1)^2 -d_0\gamma_1^2 - d_1\beta_1^2
           \eql d_1 \frac{l^2}{r^2}(d_1-1-\kappa^2\phii^2).
\ear

    Now, it can be easily verified that one should necessarily put $\beta_1
    =1$. Indeed, for any $\beta_1\ne 0$ we have $r = \e^{\beta} \sim
    l^{\beta_1}$. So $\beta_1 <1$ is excluded since it leads to $r\ll l$,
    contrary to the above requirement. But if we suppose $\beta_1 > 1$, then
    $l^2/r^2 \to 0$ as $l\to\infty$, and \eq (\ref{int-A2}) leads either (if
    $\gamma_1=0$) to $\beta_1^2 =0$  or (if $\gamma_1 \ne 0$ and then
    $d_0\gamma_1 + d_1\beta_1 =1$) to $d_0\gamma_1^2 + d_1\beta_1^2 =1$.
    Both opportunities contradict the assumption $\beta_1 > 1$.

    Thus $\beta_1 = 1$, hence $r \approx kl$, $k = \const >0$, at large $l$.

    \eq (\ref{00-A2}) now leaves two opportunities, $\gamma_1 =0$ and
    $\gamma_1 = -(d_1 - 1)/d_0$, and we consider them separately in items
    (a) and (b). Item (c) describes a case when the expansions (\ref{l-}) do
    not work.

\medskip\noi
    {\bf (a)} $\gamma_1 = 0$. Then \eq (\ref{22-A2}) yields
\beq
    k^2 = 1 - \frac{\kappa^2\phii^2}{d_1-1},             \label{k_a}
\eeq
    and \eq (\ref{int-g}) in the same order of magnitude is satisfied
    automatically. The metric takes the following asymptotic form:
\beq
      ds^2 = \e^{2\gamma_{\infty}} \eta\mn dx^\mu dx^\nu     \label{ds-A2a}
                        -dl^2 - k^2 l^2 \,d\Omega^2,
\eeq
    where $\gamma_{\infty}$ is a constant (we cannot turn it to zero by
    rescaling the coordinates $x^\mu$ since such an operation has already
    been done for making $\gamma=0$ at the center).

    Thus the whole metric has a flat asymptotic, up to a solid angle deficit
    in the spherical part due to $k^2 \ne 1$. Such a deficit is a common
    feature of topological defects in cases when they have (almost) flat
    asymptotics. Its appearance due to cosmic strings and global monopoles
    in space-times without extra dimensions is discussed in detail in
    \cite{Vilenkin Shellard}. For a global monopole in extra dimensions in
    the particular case $d_0=4,$ $d_1=2$ it was treated by Benson and
    Cho \cite{Benson Cho}. We would like to stress that the situation of a
    quasi-flat asymptotic with a solid angle deficit is not general.
    It takes place only for potentials with zero value at the minimum,
    $V(\phii) =0$, and even in this case not always --- see item B.
    Namely, this geometry requires
\beq                                                        \label{phi_cr}
    |\phii| < \phi_{\rm cr} := \sqrt{d_0-1}/\kappa,
\eeq
    i.e., $\phii$ should be smaller than the critical value $\phicr$ related
    to the $D$-dimensional Planck length. As $\phii$ approaches $\phicr$,
    $k \to 0$, the deficit absorbs the whole solid angle, and the above
    geometry disappears.

    The scalar equation (\ref{phi-g}) shows how $\phi$ approaches $\phii$:
    in the senior order, it reads
\beq                                                        \label{phi-A2}
       - \frac{d_1\phii} {k^2 l^2}
                = V_{\phi\phi} (\phii)\cdot (\phi-\phii).
\eeq
    Assuming $V_{\phi\phi}(\phii) \ne 0$, we obtain $\phi-\phii \sim 1/l^2$.

\medskip\noi
    {\bf (b)} $\gamma_1 = -(d_1-1)/d_0$. Now \eq (\ref{22-A2}) leads to
    $\kappa^2 \phii^2 = d_1-1$, i.e., $\phii = \phicr$, while a substitution
    to (\ref{int-g}) gives $(d_1-1)(d_0+d_1-1) = 0$, contrary to our
    assumption $d_1-1 > 0$. So this opportunity does not lead to a regular
    asymptotic.

\medskip\noi
    {\bf (c)} If $V(0) = 0$, then a regular integral curve, starting at
    $l = 0$ and $\phi = 0$, can finish again with $\phi \to 0$ as
    $l \to \infty$. Assuming large $l$ and $r$, the scalar field equation
    (\ref{phi-g}) for $|\phi| \ll 1$ reduces to
\beq                                                       \label{phi-A2c}
    \phi''+ (d_0 \gamma' + d_1 \beta')\phi' - V_2 \phi = 0,
\eeq
    where $V_2 = V_{\phi\phi}(0)$. Since, by assumption, $\phi=0$ is a
    maximum of $V(\phi)$, we assume $V_2 < 0$.

    If we further assume that the function $s(l) = \e^{d_0\gamma + d_1\beta}$
    satisfies the condition $s''/s \to 0$ as $l \to\infty$ (which is the
    case, e.g., for any power-behaved function), the solution to
    \eq (\ref{phi-A2c}) is an oscillating function at large $l$:
\beq
   \phi \approx \phi_0 \e^{-(d_0\gamma + d_1\beta)/2}
    \cos\left [\sqrt{|V_2|} (l-l_0) \right],\cm   l \to \infty,
\eeq
   where $\phi_0$ and $l_0$ are arbitrary constants. Substituting this to
   \eq (\ref{00-g}) and averaging $(\cos)^2 \to 1/2$, we get
\beq                                                        \label{gam-A2c}
    \e^{d_0\gamma} \approx \frac{d_0 \kappa^2 |V_2|
            \phi_0^2}{2(D-2)}
             \int^{l} \frac {l\,dl} {r^{d_1}},\cm  l\to\infty.
\eeq
   It is easy to verify that for $d_1 > 2$, when the integral in
   (\ref{gam-A2c}) converges, the asymptotic form of the solution for
   $r=\e^{\beta}$ and $\gamma$ is $r \approx l$ and
\[
    \gamma = \gamma_{\infty} - \gamma_1/l^{d_{1}-2},\cm
                \gamma_1,\ \gamma_{\infty} = \const,
\]
   i.e., we have a flat asymptotic.

   In the special case $d_{1}=2$, the integral diverges logarithmically, and
   the solution may be approximated as (again) $r \approx l$ and
   $\e^{\gamma} \approx \const \cdot\ln l$.  This ``logarithmic'' asymptotic
   resembles the behavior of cylindrically symmetric solutions in standard
   \GR.

\subsection* {B: Solutions with the asymptotic $r\to r_* > 0$}

    A regular solution evidently cannot terminate at finite $r$ and
    $l < \infty$. Therefore we seek regular asymptotic $l \to \infty$,
    where $r$ and $\phi$ tend to finite limits, $r_*$ and $\phi_*$, hence
    the quantities $\beta',\ \beta'',\ \phi',\ \phi''$ vanish.

    Moreover, in a regular solution, $\gamma'$ should tend to a finite limit
    as $l \to \infty$, hence $\gamma''\to 0$. As a result, \eqs
    (\ref{00-g}) and (\ref{22-g}) at large $l$ lead to
\beq
     d_0 \gamma'^2 = - \oV_*                                   \label{B}
                    = \frac{1}{r_*^2}(\kappa^2\phi_*^2 -d_1 + 1),
\eeq
    where $\oV_* = \oV (\phi_*)$. We see that $\oV_* \leq 0$ and, in
    addition, the scalar field should be critical or larger,
    $\phi_* \geq \phicr$. According to (\ref{B}), at large $l$
\beq
    \pm \gamma' \approx h := \sqrt{-\oV_* /d_0} \geq 0,    \label{def-h}
\eeq
    and \eq (\ref{int-g}), as in the previous cases, simply verifies that
    the solution is correct in the leading order of magnitude. The scalar
    field equation gives a finite asymptotic value of $V_\phi\equiv dV/d\phi$:
\beq
    V_\phi (\phi_*) = -d_1 \phi_* r_*^{-2}.                 \label{slope-B}
\eeq
    This value is negative if $\phi_* > 0$.

    We obtain different asymptotic regimes for negative, positive and zero
    values of $\gamma'$.

\medskip\noi
    {\bf B1:} $\e^\gamma \sim \e^{-h l}$, $h > 0$. The metric has the
    asymptotic form
\beq                                                            \label{ds-B1}
      ds^2 = C^2\e^{-2hl} \eta\mn dx^\mu dx^\nu - dl^2 - r_*^2 d\Omega^2.
\eeq
    The extra-dimensional part of the metric describes an infinitely
    long cylindrical tube, but the vanishing function
    $g_{tt}=\e^{2\gamma}$ resembles a horizon. The substitution $\e^{-h l} =
    \rho$ (converting $l=\infty$ to a finite coordinate value, $\rho=0$)
    brings the metric (\ref{ds-B1}) to the form
\beq                                                           \label{hor-B1}
      ds^2 = C^2 \rho^2 \eta\mn dx^\mu dx^\nu
                   - \frac{d\rho^2}{h^2 \rho^2} - r_*^2 d\Omega^2.
\eeq
    Thus $\rho=0$ is a second-order Killing horizon in the
    2-dimensional subspace parameterized by $t$ and $\rho$, it is of the same
    nature as, e.g., the extreme \RN\ \bh\ horizon, or the \AdS\ horizon in
    the second Randall-Sundrum \bw\ model. A peculiarity of the present
    horizon is that the spatial part of the metric, which at large $l$ takes
    the form $\rho^2 (d\vec x)^2$, is {\it degenerate\/} at $\rho=0$.
    The volume of the $d_0$-dimensional spacetime vanishes as $l \to
    \infty$. And it will remain degenerate even if we pass to Kruskal-like
    coordinates in the ($t,\rho$) subspace. But the $D$-dimensional
    curvature is finite there, indicating that a transition to negative
    values of $\rho$ (where the old coordinate $l$ no longer works) is
    meaningful.
\footnote
      {One may wonder why we here do not obtain simple (first-order)
       horizons, like those in the \Sch\ and \dS\ metrics, while such
       horizons generically appeared in the special case $d_0 = 1$, which
       corresponds to \sph\ global monopoles in \GR, considered in detail
       in Refs.\,{\cite{Bronnikov Global mon, Bronnikov Multi-D}}.

       The reason is that, in case $d_0 = 1$, $\delta\MN{}_{\rho\sigma}$
       (\ref{Delta*munu_rosigma}) is zero, and the corresponding component
       of the Riemann tensor is also zero regardless of the values of
       $\gamma'$. In terms of the Gaussian coordinate $l$, a simple horizon
       occurs at some finite $l = l_h$ near which $g_{tt} = \e^{2\gamma}
       \sim (l - l_h)^2$, so that $\gamma' \to \infty$. When $d_0 = 1$
       this does not lead to a singularity since then only the combinations
       $\gamma''+\gamma'{}^2$ and $\beta'\gamma'$ are required to be (and
       really are) finite. In case $d_0 > 1$, instead of a horizon, we would
       have a curvature singularity at finite $l$, a situation excluded in
       the present study.

       We thus have a {\it general\/} result for the metric (\ref{ds}):
       {\it for $d_0 > 1$, horizons can only be of order 2 and higher.}}

    One more observation can be made. According to (\ref{B}), the potential
    $V$ is necessarily negative at large $l$. On the other hand, \eq
    (\ref{00-g}) may be rewritten in an integral form:
\beq
      \e^{d_0\gamma + d_1\beta} \gamma'                     \label{00-int}
              = - \int_0^l  \e^{d_0\gamma + d_1\beta}\, \oV\, dl.
\eeq
    The lower limit of the integral corresponds to a regular center, where
    the left-hand side of (\ref{00-int}) vanishes. As $l\to\infty$,
    it vanishes as well due to $\gamma \to -\infty$. Thus the integral in
    the right-hand side, taken from zero to infinity, is zero. It means
    that the potential $V(\phi)$ has an alternate sign and is positive in a
    certain part of the range $(0,\ \phi_*)$.

    Thus purely scalar solutions of monopole type may contain second-order
    horizons. The degenerate nature of the spatial metric at the horizon
    does not lead to a curvature singularity, and the solutions may be
    continued in a Kruskal-like manner. Nevertheless, we do not consider
    these solutions as describing viable monopole configurations because the
    zero volume of the corresponding spatial section will make the density
    of any additional (test) matter become infinite. It is then impossible
    to neglect its back reaction, which will evidently destroy such a
    configuration.

\medskip\noi
    {\bf B2:} $\e^\gamma \sim \e^{h l}$, $h > 0$. The metric has the
    asymptotic form
\beq                                                            \label{ds-B2}
      ds^2 = C^2\e^{2hl} \eta\mn dx^\mu dx^\nu - dl^2 - r_*^2 d\Omega^2,
            \cm\cm C = \const > 0.
\eeq
    So, in the \sph\ extra-dimensional part of the metric, we
    have an infinitely long $d_1$-dimensional cylindrical ``tube'' with an
    infinitely growing gravitational potential $g_{tt} = \e^{2\gamma}$.

    At this cylindrical asymptotic, according to (\ref{def-h})
    and (\ref{slope-B}), the potential $V$ tends to a negative value and has
    a negative slope. Moreover, in \eq (\ref{00-int}), the integral
    is negative and diverges at large $l$ due to growing $\e^{\gamma}$.

    Regular solutions with $\gamma'(\infty) > 0$ naturally arise if the
    potential $V(\phi)$ is negative everywhere. One can notice, however,
    that when $V(0)$ is above zero, by (\ref{00-int}), near the center
    ($l=0$) the function $\gamma(l)$ decreases due to $V > 0$ while at large
    $l$ it grows. So it has a minimum at some $l>0$.

\medskip\noi
    {\bf B3:} $\oV_* = 0$. This case contains one more asymptotic where the
    extra space ends with a regular tube.

    Indeed, we can once again use the expansions (\ref{l-}), but now with
    $\phi_*$ instead of $\phii$ and $\beta_1 = 0$ in accord with $r\to r_*$.
    \eq (\ref{22-g}) [order $O(1)$] shows that $\kappa^2 \phi_*^2 = d_1-1$,
    i.e., $\phi_* = \phicr$. \eq (\ref{phi-g}) [order $O(1)$] gives a finite
    value of the derivative $dV/d\phi (\phi_*) = -d_1 \phi_*/ r_*^2$.
    Further, \eq (\ref{00-g}) [$O(l^{-2}$)] yields
    $\gamma_1 (d_0\gamma_1 - 1)/l^2 = -\oV$, showing that $V = O(l^{-2})$
    (or is even smaller). Since $V = dV/d\phi(\phi_*)\cdot (\phi-\phi_*)
    + o(\phi-\phi_*)$, we have to conclude that $\phi-\phi_* = O(l^{-2})$ or
    smaller.

    Now, assuming $V(\phi) = V_2/l^2 + \ldots$, we can find $V_2$ directly
    as the senior term in  $dV/d\phi(\phi_*)\cdot (\phi-\phi_*)$ and,
    independently, from \eq (\ref{22-g}) [$O(l^{-2})$], getting the following
    two expressions:
\[
      V_2 = - d_1 \frac{\phi_*\phi_2}{r_*^2}
                \cm {\rm and} \cm
      V_2 = - (D-2) \frac{\phi_*\phi_2}{r_*^2},
\]
    whence it follows $d_1 = D-2$, or $d_0 = 1$. Such a ``critical''
    asymptotic ($\phi \to \phicr$, $g_{tt} \to 0$, $r\to \const$) was indeed
    found for $d_0=1$ in our papers \cite{Bronnikov Global mon, Bronnikov
    Multi-D} describing $(d_1+2)$-dimensional \sph\ global monopoles, but,
    as we see, it does not exist in the case under consideration, $d_0 > 1$.

    The only remaining opportunity is that $\phi - \phi_* = o(l^{-2})$ and
    $\gamma \to \gamma_* = \const$, i.e., a solution tending at large $l$ to
    the following simple ``flux-tube'' solution, valid for any
    $d_0$ and $d_1$:
\beq                                                          \label {tube}
      r = \const, \cm \gamma = \const, \cm
                 \phi= \phicr, \cm V =0, \cm dV/d\phi = -d_1 \phicr/r^2.
\eeq
    Such a solution can exist if the potential $V(\phi)$ has the properties
    $V(\phicr) = 0$ and $dV/d\phi (\phicr) < 0$, and the last equality
    in (\ref{tube}) then relates the constant radius $r$ to $dV/d\phi
    (\phicr)$.

\subsection* {C: Solutions with the asymptotic $r\to 0$}

    The limit $r\to 0$ means a center, and for it to be regular, the
    conditions (\ref{reg-g}) should hold, hence, for our system, the
    initial conditions (\ref{cond_0}), where $l=0$ should be replaced with,
    say, $l=l_0 > 0$.

    Recall now that the conditions (\ref{cond_0}) determine the solution to
    the field equations for a given potential $V(\phi)$ up to the value of
    $\phi'$. In particular, if there is one more center at $l = l_0$, then,
    starting from it and choosing $\phi'(l_0) = - \phi'(0)$, we shall have
    {\it the same\/} solution in terms of $l_0 -l$ instead of $l$. We thus
    obtain a solution with two regular centers which is {\it symmetric\/}
    with respect to the middle point $l = l_0/2$, to be called the {\it
    equator\/}. To be smooth there, it should satisfy the conditions
\beq
      \beta' = \gamma' = \phi'=0 \cm {\rm at} \quad l = l_0/2, \label{l_0/2}
\eeq
    which implicitly restrict the shape of the potential. Given a potential
    $V(\phi)$, the conditions (\ref{l_0/2}) create, in general, three
    relations among $l_0$, $\phi'(0)$ and the free parameters of $V(\phi)$
    (if any).  Excluding $l_0$ and $\phi(0)$, we must obtain a single ``fine
    tuning'' condition for the parameters of the potential.

    A necessary condition for the existence of such a solution is that
    $V(\phi)$ should have a variable sign. This follows from \eq
    (\ref{00-int}) with integration over the segment $(0, l_0/2)$: the
    integral vanishes since $\gamma' = 0$ at both ends.

    Moreover, as follows from \eqs (\ref{22-g}) and (\ref{int-g}) with
    (\ref{l_0/2}),
\bear                                                           \label{equa}
     r_e^{-2} (d_1 - 1 - \kappa^2\phi_e^2) = \frac{D-2}{d_1} \oV_e
                           = \beta''_e + \oV_e,
\ear
    leading to $d_1\beta''_e = (d_0-1)\oV_e$ (where the index ``e'' refers
    to values at the equator). If $r=\e^{\beta}$ is assumed to grow
    monotonically from zero to its maximum value at the equator, we have
    $\beta''_e < 0$, hence $\oV_e < 0$, and (\ref{equa}) implies that
    $\phi_e > \phicr$, i.e., the scalar field at the equator should exceed
    its critical value.

    The existence of asymmetric solutions with two regular centers,
    corresponding to $\phi'(l_0) \ne -\phi'(0)$, is also not excluded.
    In this case, there would be, in general, no equator since $\beta$ and
    $\phi$ would have maxima at different $l$; moreover, there would be,
    in general, $\gamma(l_0) \ne \gamma(0) = 0$, and $\gamma(l)$ could even
    have no extremum. However, since $\gamma' =0$ at both centers,
    the integral in (\ref{00-int}) taken from 0 to $l_0$ should vanish,
    hence, again, $V$ would have an alternating sign.

    The whole configuration with two regular centers has the topology
    $\M^{d_0} \times \S^{d_1 + 1}$, with closed extra dimensions in the
    spirit of Kaluza-Klein models. The main difference from them is that now
    all variables essentially depend on the extra coordinate $l$.

    The main properties of all regular asymptotics found, which lead to a
    classification of possible global monopole configurations in extra
    dimensions, are summarized in Table 2. The word ``attraction''
    corresponds to an increasing warp factor far from the brane.

\begin{table}[bh]
\caption{Classification of global monopole solutions for arbitrary $V(\phi)$
     by asymptotic types. Attraction or repulsion is meant with respect to
     the center.}
\begin{center}              \tabcolsep 10pt
\begin{tabular}{|c|c|c|c|c|l|}
\hline
 Notation   &  $r$  &  $V(\phi)$   &  $\phi$ & $\gamma$  &
                                Asymptotic type $\wide$\\
\hline
$\wide$ A1  & $\infty$ & $V(\eta) < 0$   & $\eta<\phicr$ & $\infty$
                        & AdS, attraction         \\
 A2(a) & $\infty$ &  0 & $\eta<\phicr$ & const & flat, solid angle deficit\\
 A2(c), $d_1>2$ & $\infty$ &  0 &    0    & const    & flat               \\
 A2(c), $d_1=2$ & $\infty$ &  0 &    0    & $\infty$ & ``logarithmic'',
                             attraction       \\
       &&&&&\\[-10pt]
\hline
       &&&&&\\[-10pt]
 B1 & $r_*$ & $V_* <0$ & $\phi_* >\phicr$ & $-\infty$ & double horizon,
                             repulsion \\
 B2 & $r_*$ & $V_* <0$ & $\phi_* >\phicr$ &  $\infty$ & attracting tube \\
 B3 & $r_*$ &   0      & $\phi_* =\phicr$ &  const    & trivial tube   \\
       &&&&&\\[-10pt]
\hline
 C  &   0   & $V(0)$   &   0              &  const & second center $\wide$\\
\hline
\end{tabular}
\end{center}
\end{table}

\section {Scalar field trapping by global monopoles}  

    Consider a test scalar field with the Lagrangian
    (\ref{L_hi}) in the background of global monopole configurations
    described in \sect 4. After variable separation (\ref{sepa}), the
    field equation for a $\vp$-mode of the scalar field $\chi$ reads
\beq
    X'' + (d_0 \gamma' + d_1 \beta')X'                    \label{X''D}
                + (\e^{-2\gamma} p^2 -\mu^2) X =0,
\eeq
    where $p^2 = p_\mu p^\mu = E^2 - \vp^2$ is the $d_0$-momentum squared
    and $\mu^2 = m_0^2 + \lambda \phi^2$ is the effective mass squared. The
    trapping criterion consists, as before, in the requirements that
    the $\chi$ field SET should vanish far from the brane and the
    total $\chi$ field energy per unit volume of the brane should be finite,
    i.e.,
\bear                                                        \label{E_D}
     E_{\rm tot}[\chi] = \int \sqrt{g} d^{d_1+1}x
            \Bigl[ \e^{-2\gamma}(E^2 + \vp^2)X^2
                        + \mu^2 X^2 + X'^2 \Bigr] dl < \infty.
\ear
    The first requirement means that each term in the square brackets in
    (\ref{E_D}) must vanish at large $l$.

    Let us now check whether these requirements can be met at different
    kinds of asymptotics listed in Table 1.

\medskip\noi
    {\bf A1:} attracting AdS asymptotic, $\beta \sim \gamma \sim kl$, $k>0$.
    At large $l$, \eq (\ref{X''D}) reduces to the equation with constant
    coefficients $X'' + (D-1)X - \mu^2 X =0$, and its solution vanishing as
    $l\to \infty$ is
\beq                                                         \label{X-A1}
    X \sim \e^{-al}, \cm
        a = \half \left[(D-1)k + \sqrt{(D-1)^2 k^2 + 4\mu^2}\right].
\eeq
    It is straightforward to make sure that the trapping requirements are
    satisfied for all momenta $\vp$ and all $\mu^2 \geq 0$.

\medskip\noi
    {\bf A2(a):} a quasi-flat asymptotic with a solid angle deficit.
    At large $l$, \eq (\ref{X''}) reduces to $X''+d_1 X/l + P^2 X =0$,
    where $P^2 = p^2 \e^{-2\gamma_\infty} - \mu^2$ and $\gamma_\infty$ is
    the limiting value of $\gamma$ at $l=\infty$. In terms of $Y=
    l^{d_1/2}X$ this equation is (at large $l$) rewritten as $Y'' + P^2 Y
    =0$, while the trapping condition (\ref{E_D}) implies
    $\int l^{d_1} X^2(l)\, dl < \infty$. Therefore only an exponentially
    falling $Y(l)$ is suitable. In other words, the trapping condition is
    $P^2 < 0$, or
\beq
    p^2 < m_{\rm cr}^2 := \mu^2\e^{2\gamma_{\infty}},      \label{p-A2}
\eeq
    where now $\mu^2 = m_0^2 + \lambda^2 \eta^2$. One should note that $p^2
    = E^2 - \vp^2$ is nothing else but the observable mass of a free
    $\chi$-particle if the observer watches its motion in the Minkowski
    section $l=0$ of our manifold, i.e., on the brane. So the condition
    (\ref{p-A2}) means that {\it the brane traps all scalar particles of
    masses smaller than the critical value $m_{\rm cr}$\/} depending on the
    model parameters.

\medskip\noi
    {\bf A2(c), $d_1 > 2$:} this case differs from the previous one only by
    the asymptotic value of $\phi$ which is now zero, so that $\mu = m_0$.

\medskip\noi
    {\bf A2(c), $d_1=2$:} a ``logarithmic'' asymptotic, $\e^\gamma\sim \ln l$.
    Since $\e^{-2\gamma} \sim 1/(\ln l)^2 \to 0$, the term with $p^2$ drops
    out from \eq (\ref{X''D}), which then leads to the decreasing
    solution $X\sim l^{-1}\e^{-\mu l}$, and a $\chi$-particle is trapped
    provided $\mu = m_0 >0$.

\medskip\noi
    {\bf B1:} a horizon. As was remarked previously, we do not regard this
    configuration viable and omit it in our discussion.

\medskip\noi
    {\bf B2:} an attracting tube, $r\to r_*$ and $\gamma \approx hl$, $h>0$
    as $l\to \infty$. \eq (\ref{X''D}) takes the form
    $X'' + d_0 h X' - \mu^2 X =0$ and has the decreasing solutions
\beq
      X \sim \e^{-al}, \cm                                     \label{X-B2}
            a = \half \left(d_0 h + \sqrt{d_0^2 h^2 + 4 \mu^2}\right).
\eeq
    As in item A1, it is easy to verify that the trapping conditions hold
    provided $\mu^2 >0$.

\medskip\noi
    {\bf B3:} a trivial tube, both $\beta$ and $\gamma$ tend to constants as
    $l \to \infty$. In \eq (\ref{X''D}) the term with $X'$ drops out at
    large $l$, and an exponentially decreasing solution exists under the
    condition (\ref{p-A2}) where $\mu^2 = m_0^2 + \lambda^2 \phicr^2$.

\medskip\noi
    {\bf C:} these configurations have no large $l$ asymptotic and are not
    interpreted in terms of branes.

\medskip
    A conclusion is that scalar particles of any mass and momentum are
    trapped by global monopoles with A1 and B2 asymptotics with
    exponentially growing warp factors and A2(c) with a logarithmic
    asymptotic; they are trapped under the restrictions (\ref{p-A2}) on the
    particle's observable mass by monopoles with A2 and B3 asymptotics whose
    warp factors tend to constant limits far from the brane.

\section {Numerical results: Mexican hat potential} 

    In this section we present the results of our numerical calculations
    which confirm the classification of regular solutions given above. In
    doing so, we have been using the ``Mexican hat'' potential in the form
    (Fig.\,1)
\beq
    V=\frac{\lambda \eta^4}{4} \biggl[ \eps +\biggl( 1-\frac{\phi^{2}}
    {\eta^{2}} \biggr)^{2} \biggr].                       \label{V_Mex}
\eeq
\begin{figure}[h]
\hspace{6cm}
\includegraphics{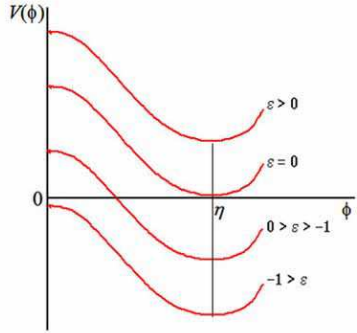}
\caption{Mexican hat potential}
\end{figure}
    It has two extremum points in the range $\phi\geq 0$: a maximum at $\phi
    =0$ and a minimum at $\phi =\eta$. The SSB energy scale is characterized
    by $\eta^{1/(D-2)}$, while $\sqrt{\lambda}\eta $ determines, as usual, a
    length scale. The non-conventional parameter $\eps$, introduced in
    (\ref{V_Mex}), moves the potential up and down, which is equivalent to
    adding a cosmological constant to the usual Mexican hat potential.

    Given the potential (\ref{V_Mex}), the nature of the solutions
    essentially depends on its two dimensionless parameters: $\eps$, fixing
    the extremal values of the potential with respect to zero, and
    $\kappa^2\eta^2$ characterizing the gravitational field strength: as we
    remember from \sect 4, the asymptotic $r\to\infty$ only exists when
    $\phii < \phicr$, which is the same as $\kappa^2\eta^2 < d_1-1$.

    If $\eps > 0$, the potential (\ref{V_Mex}) is always positive, and, in
    accordance with \sect A1, regular solutions are absent.

    In the conventional case $\eps =0$, in the range $0 < \kappa^2\eta^2
    < d_1 - 1$, there are asymptotically flat regular solutions with a solid
    angle deficit (class A2).

    The most complex case $0 > \eps > -1$ contains a variety of
    possibilities. Regular solutions with the asymptotic $r\to \infty $ as
    $l\to \infty $ having $\gamma'_{\infty}>0$ (case A1) exist in some
    range $0 < \eta < \eta_s$, where the separating value $\eta_s$
    depends on $d_0,\ d_1$ and $\eps$.  As an example, such a regular
    solution with $\kappa^2 \eta^{2} = 5$, $\eps = -0.75$, $d_{0} = 4$,
    $d_{1} = 3$ is presented in Fig.\,2.

\begin{figure}[h]
\includegraphics{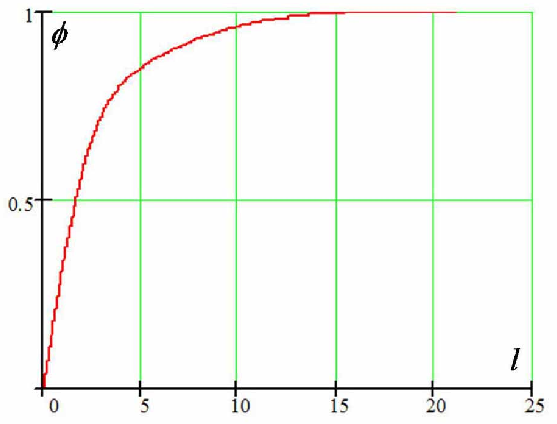}
\includegraphics{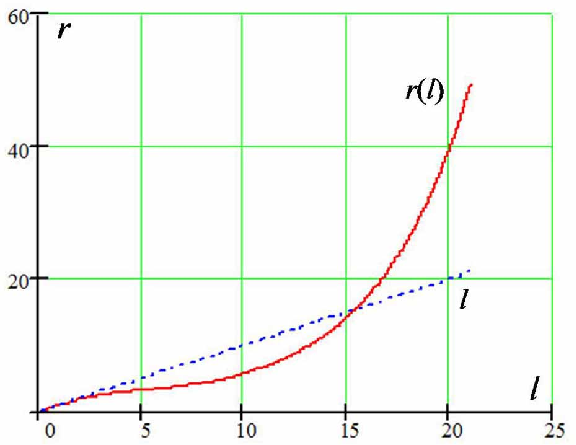}
\includegraphics{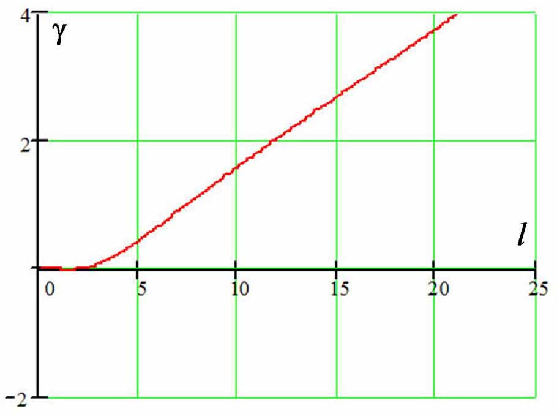}
\caption{A regular solution with an AdS asymptotic (type A1) for the
    potential (\ref{V_Mex}) with $\kappa \eta ^{2}=5$, $\eps =-0.75$,
    $d_{0}=4$, $d_{1}=3$}
\end{figure}
    Depending on the parameters of the potential, there are regular
    solutions with the asymptotic $r\to r_* < \infty $ and
    $\gamma'_{\infty} > 0$  (case B2) in
    some range $\eta_{s1} < \eta <\eta_{s2}$, see Fig.\,3.

\begin{figure}[h]
\includegraphics{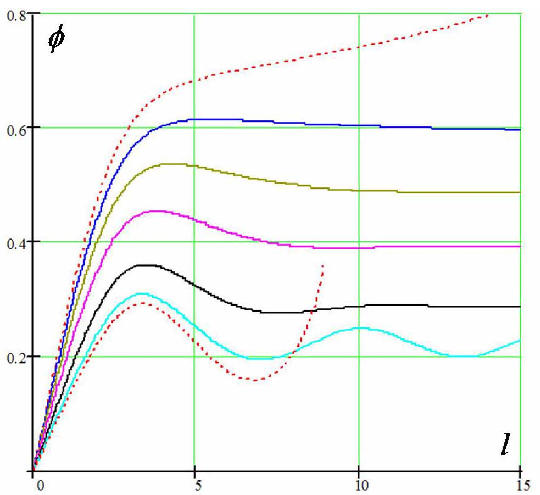} 
\includegraphics{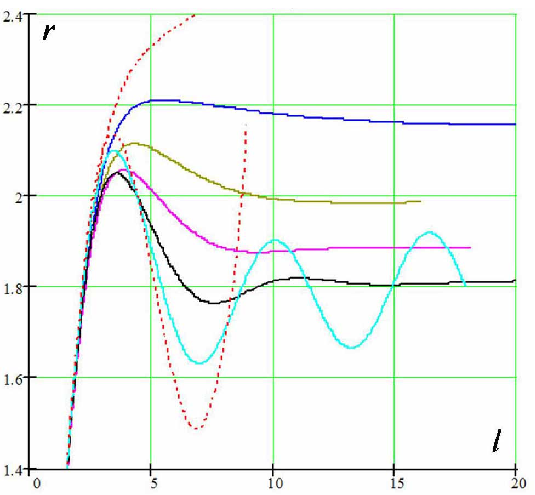} 
\includegraphics{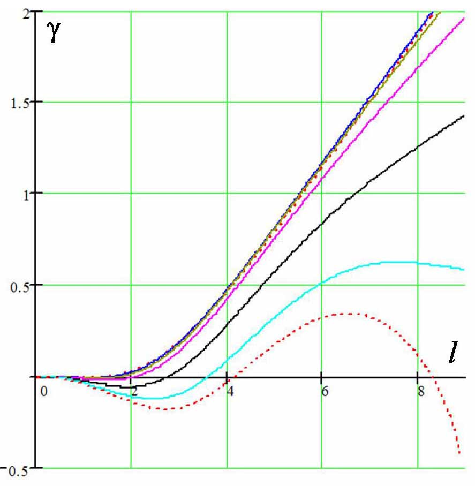}
\caption{Regular solutions with the asymptotic B2 (attracting tube),
    such that $r\to r_*<\infty $ and $\gamma'_{\infty } > 0$}
\end{figure}
    Here $\eps =-0.9$, $d_{0}=4$, $d_{1}=3$. The curves are given for
    $\kappa^2 \eta^2=$ 10, 12, 15, 20, 30, 40 and 45 (upside down).
    The red dotted curves ($\kappa^2 \eta^{2}=10$ and $\kappa^2 \eta^2 =
    45$) correspond to singular configurations. So for $\eps =-0.9$,
    $d_{0}=4$, $d_{1}=3$ the lower bound of this parameter leading to
    regular models is somewhere between 10 and 12 while the upper bound is
    between 30 and 45.

    An example of a regular solution with the asymptotic $r\to r_* < \infty$
    and $\gamma _{\infty }'<0$ (class B1), corresponding to a second-order
    Killing horison, is shown in Fig.\,4.
\begin{figure}[h]
\includegraphics{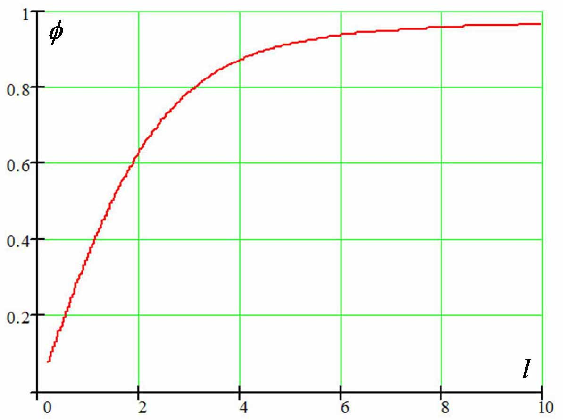} 
\includegraphics{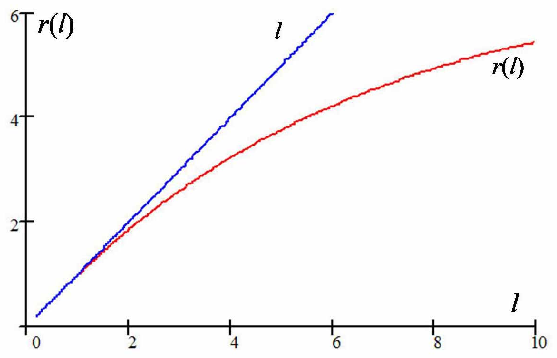} 
\includegraphics{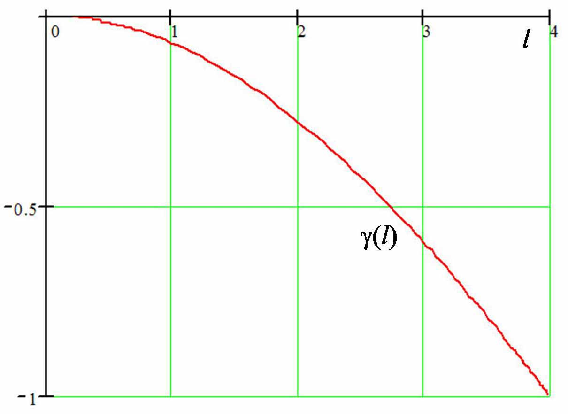}
\caption{A regular solution with the asymptotic $r\to r_*<\infty $ and
    $\gamma _{\infty }'<0$ (case B1, horizon).}
\end{figure}
    The value $\kappa \eta^{2} = 17.37$ is fine-tuned to the parameters
    $\eps =-0.75$, $d_{0}=4$, $d_{1}=2$ of this particular solution.

    Other examples of fine-tuned regular solutions, namely, type C
    with two regular centers ($r\to 0$, $\phi \to 0$,
    $\gamma'\to 0$ at $l\to l_{0})$, are presented in Fig.\,5.
\begin{figure}[h]
\includegraphics{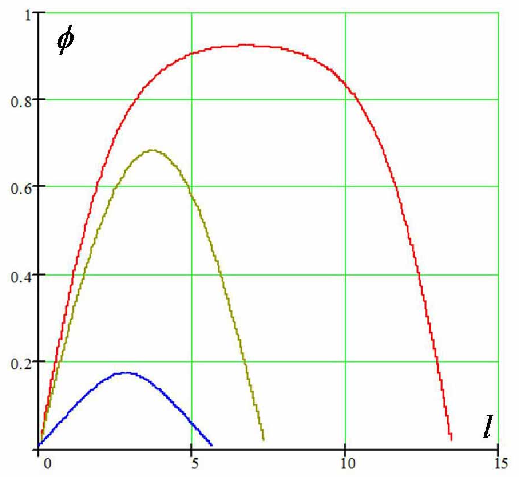} 
\includegraphics{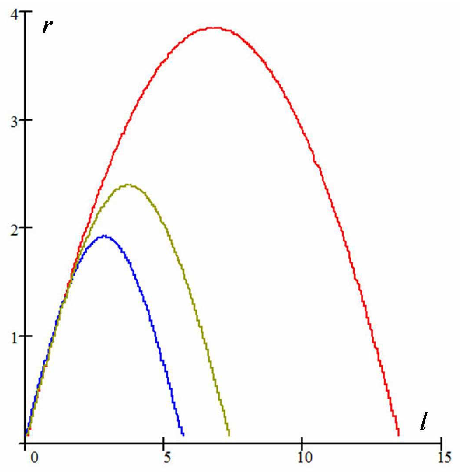} 
\includegraphics{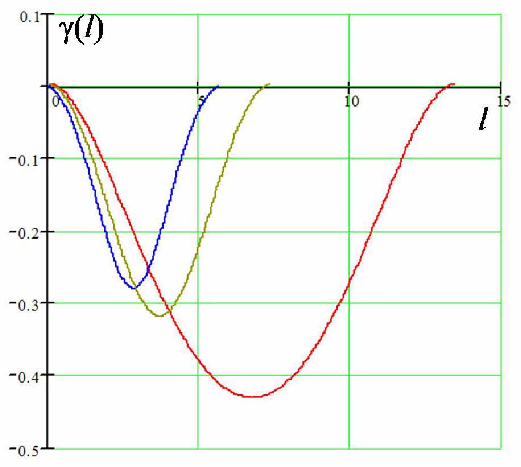}
\caption{Type C solutions with two regular centers ($r\to
    0,\ \phi \to 0,\ \gamma'\to 0$ as $l \to l_{0})$}
\end{figure}
\begin{figure}[h]
\includegraphics{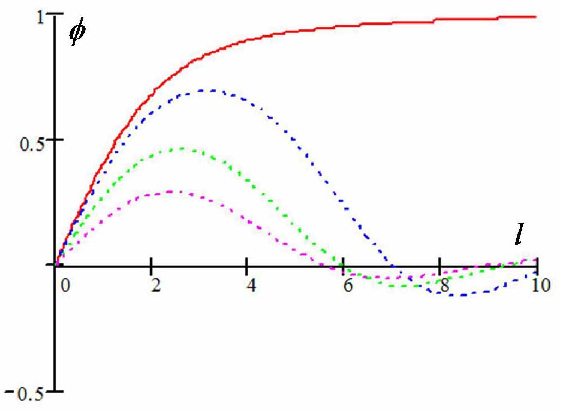} 
\includegraphics{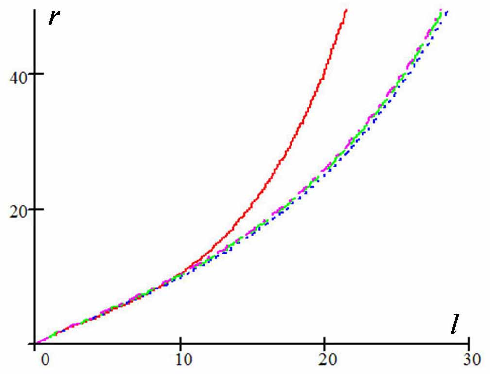} 
\includegraphics{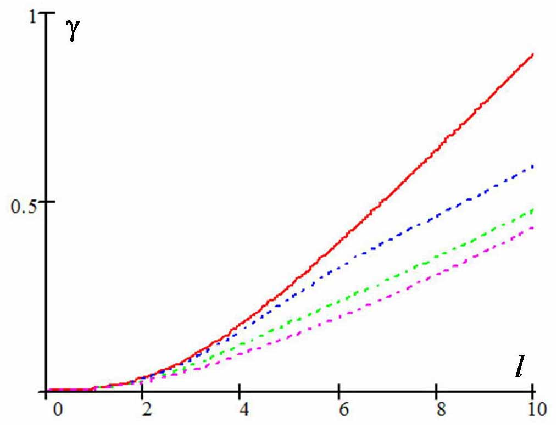}
\caption{Regular solutions starting and terminating at $\phi
    (0)=\phi (\infty) =0$. The limiting red curve with
    $\phi '(0) =\phi _{c}'=0.4401425$
    (separatrix) terminates at $\phi (\infty) =\eta$.}
\end{figure}
    For all three curves $d_{0}=4$, $d_{1}=2$. The red, green and blue
    curves correspond to $\eps =-0.15$, $-0.5$, and $-0.9626$, respectively.
    The fine-tuned values of $\kappa^2 \eta^2$ are approximately 2.637,
    6.17 and 100.

    In case $\eps \leq -1$, the maximum $V(0)\leq 0$ is at or below the zero
    level, and there is a possibility for the integral curves to start and
    finish at the same value $\phi (0) = \phi (\infty) =0$. We then observe
    a whole family of such regular curves in the range $0 < \phi'(0) <\phi'
    _{c}$, see Fig.\,6.
    For the particular example presented ($\eps =-1.5$, $\kappa \eta^2=1$,
    $d_{0}=4$, $d_{1}=3)$, the values of $\phi'(0)$ for the dotted curves
    ending with $\phi =0$ are 0.2, 0.3 and 0.4 (bottom up). The limiting red
    curve with $\phi'(0) =\phi _{c}'=0.4401425$ (separatrix) is a regular
    solution ending at the minimum of the potential: $\phi \to \eta $ as
    $l\to \infty$.

    The Mexican hat potential (\ref{V_Mex}), with its only two extrema at
    $\phi=0$ and $\phi= \eta$, cannot demonstrate the whole variety
    of solutions which appear with more sophisticated potentials having
    additional maxima and/or minima. Thus, for instance, class A solutions
    may have a large $r$ asymptotic at any such extremum.

\section {Concluding remarks}

    We have obtained as many as seven classes of regular solutions to the
    field equations describing a Minkowski thick brane with a global
    monopole in extra dimensions, see Table 1.

    Some of these classes, namely, A1 with an AdS asymptotic and B2 ending
    with an attracting tube, possess an exponentially growing warp factor
    $\e^{2\gamma}$ at large $l$ and are shown to trap linear test scalar
    fields modes of any mass and momentum.

    Others --- A2(a) and A2(c) for $d_1 > 2$, ending with a flat metric at
    large $l$ --- have a warp factor tending to a constant whose value is
    determined by the shape of the potential $V(\phi)$. They are also shown
    to trap a test scalar field but the latter's observable mass is
    restricted above by a value depending on the particular model of the
    global monopole.

    Lastly, for $d_1=2$, i.e., a three-dimensional global monopole in the
    extra dimensions, class A2(c) solutions possess a logarithmically
    growing warp factor. All test scalar field modes are trapped by this
    configuration, but the slow growth of $\gamma(l)$ probably means that
    the test field is strongly smeared over the extra dimensions.

    All such configurations, in sharp contrast to RS2-like domain walls in 5
    dimensions, are able to trap scalar matter. It is certainly necessary
    to check whether or not nonzero-spin fields are trapped as well and
    Newton's law of gravity holds on the brane in conformity with the
    experiment. We hope to consider these subjects in our future publications.

    In addition to the trapping problem, a shortcoming of RS2 type
    Minkowski branes is that they are necessarily fine-tuned. Many of the
    global monopole solutions, at least those existing in the weak gravity
    regime (class A), are free of this shortcoming and are thus better
    for thick brane model building.


    Some results and conclusions of this paper have been previously
    found in Refs.  \cite{vil, shap}. The main difference of our approach
    from theirs is their boundary condition which is, in our notation, $\phi
    = \eta$. This excludes the cases when the solution ends at a maximum
    or slope of the potential, such as, e.g., symmetric solutions with two
    regular centers. Another difference is that they consider solutions
    with an exponentially decreasing warp factor as those leading to matter
    confinement on the brane. In our view, such solutions with second-order
    horizons do not represent viable models of a brane world. We conclude
    that the present paper gives the most complete classification of all
    regular solutions for global monopoles in extra dimensions, which, even
    without gauge fields, seem to be promising as brane world models.

\end{document}